\documentclass[12pt]{article}%
\usepackage{amsmath}
\usepackage{amsfonts}
\usepackage{amssymb}
\usepackage{graphicx}%
\providecommand{\U}[1]{\protect\rule{.1in}{.1in}}
\begin{document}
\begin{titlepage}
\ \\
\begin{center}
\LARGE
{\bf
Energy Entanglement Relation\\
for Quantum Energy Teleportation
}
\end{center}
\ \\
\begin{center}
\large{
Masahiro Hotta
}\\
\ \\
\ \\
{\it
Department of Physics, Faculty of Science, Tohoku University,\\
Sendai 980-8578, Japan\\
hotta@tuhep.phys.tohoku.ac.jp
}
\end{center}
\begin{abstract}
Protocols of quantum energy teleportation (QET), while retaining causality and local energy conservation, enable the
transportation of energy from a subsystem of a many-body quantum system
to a distant subsystem
by local operations and classical communication through ground-state entanglement.
We prove two energy-entanglement inequalities for a minimal QET model.
These relations help us to gain a profound understanding of entanglement itself as a physical resource
by relating entanglement to energy as an evident physical resource.
\end{abstract}
\end{titlepage}

\bigskip

\section{Introduction}

\bigskip

~

The relationship between energy and information has been investigated
extensively in the context of computation energy cost including a modern
analysis of Maxwell's demon \cite{L}-\cite{D}. In this Letter, we show a new
energy-information relation from a different point of view. Recently, it has
been reported that energy can be transported by local operations and classical
communication while retaining local energy conservation and without breaking
causality \cite{hotta1}-\cite{hotta3}. Such protocols are called quantum
energy teleportation (QET) and are based on ground-state entanglement of
many-body quantum systems including spin chains \cite{hotta1}, cold trapped
ions \cite{hotta2}\ and quantum fields \cite{hotta3}. By performing a local
measurement on a subsystem A of a many-body system in the ground state,
information about the quantum fluctuation of A can be extracted. Because the
post-measurement state is not the ground state in general, some amount of
energy is infused into A as QET energy input during this measurement, and the
ground-state entanglement gets partially broken. Next, the measurement result
is announced to another subsystem B of the many-body system at a speed much
faster than the diffusion velocity of the energy infused by the measurement.
Soon after the information arrives at B, energy can be extracted from B as QET
energy output by performing a local operation on B dependent on the announced
measurement data. The root of the protocols is a correlation between the
measurement information of A and the quantum fluctuation of B via the
ground-state entanglement. Due to the correlation, we are able to estimate the
quantum fluctuation of B based on the announced information from A and devise
a strategy to control the fluctuation of B. By the above-mentioned selected
local operation on B, the fluctuation of B can be more suppressed than that of
the ground state, yielding negative energy density around B in the many-body
system. The concept of negative energy density has been investigated in
quantum field theory for a long time \cite{BD}. Quantum interference among
total energy eigenstates can produce various states containing regions of
negative energy density, although the total energy remains nonnegative. The
regions of negative energy density can appear in general many-body quantum
systems by fixing the origin of the energy density such that the expectational
value vanishes for the ground state. In spite of the emergence of negative
energy density, \ the total energy also remains nonnegative for the general
cases. In the QET protocols, during the generation of negative energy density
at B, surplus positive energy is transferred from B to external systems and
can be harnessed as the QET output energy. Here it should be emphasized that
this output energy existed not at A but at B \textit{even} \textit{before} the
start of the protocol and was hidden inside the zero-point fluctuation of B.
Of course, this zero-point energy is not available by usual local operations
for B. However, by using a local operation dependent on A's information, it
becomes possible to dig out B's zero-point energy by pair creation of the
positive output energy from B and the negative energy of B. Hence, we do not
need to hire any physical carrier of energy from A to B like electric currents
and photons, at least, during short-time QET processes. Needless to say, after
the completion of QET process, the positive energy of A compensates for the
negative energy of B during late-time free evolution of the many-body system.
The amount of output energy from B is upper bounded by the amount of input
energy to A.

Using the usual protocols of quantum teleportation, quantum states can be
transported from one party to another by the consumption of shared
entanglement between the two parties \cite{qt}. As is well known \cite{nc},
transfer of a large number of quantum states requires a large amount of
consumption of shared entanglement as a physical resource. Taking into account
the fact, it seems natural for the QET protocols to expect that a large amount
of teleported energy also requests a large amount of consumption of the
ground-state entanglement between A and B. If such a non-trivial relation
exists between teleported energy and breaking of ground-state entanglement by
measurement, the relation may shed new light on the interplay between quantum
physics and quantum information theory. In this Letter, the first example of
the energy-entanglement relation for a minimal QET model is presented. The
minimal QET model is the smallest physical system for which non-trivial QET
can be implemented; this model consists of two qubits with an interaction of
the Ising spin chain in the presence of a transverse magnetic field. We
explicitly show that for the minimal model, the consumption of entanglement
between A and B during the measurement of A is lower bounded by a positive
value that is proportional to the maximum amount of energy teleported from A
to B. In addition, we obtain another inequality in which the maximum amount of
energy teleported from A to B is lower bounded by a different positive value
that is proportional to the amount of entanglement breaking between A and B by
the measurement of A. These energy-entanglement inequalities are of importance
because they help in gaining a profound understanding of entanglement itself
as a physical resource by relating entanglement to energy as an evident
physical resource.~

\section{Minimal QET Model}

\bigskip

~

First of all, we introduce the minimal QET model. The system consists of two
qubits A and B. Its Hamiltonian is the same as that of the Ising spin chain in
the presence of a transverse magnetic field as follows: $H=H_{A}+H_{B}+V$,
where each contribution is given by
\begin{align}
H_{A} &  =h\sigma_{A}^{z}+\frac{h^{2}}{\sqrt{h^{2}+k^{2}}},\label{1}\\
H_{B} &  =h\sigma_{B}^{z}+\frac{h^{2}}{\sqrt{h^{2}+k^{2}}},\\
V &  =2k\sigma_{A}^{x}\sigma_{B}^{x}+\frac{2k^{2}}{\sqrt{h^{2}+k^{2}}%
},\label{3}%
\end{align}
and $h$ and$~k$ are positive constants with energy dimensions, $\sigma_{A}%
^{x}~\left(  \sigma_{B}^{x}\right)  $ is the x-component of the Pauli
operators for the qubit A (B), and $\sigma_{A}^{z}~\left(  \sigma_{B}%
^{z}\right)  $ is the z-component for the qubit A (B). The constant terms in
Eqs. (\ref{1})-(\ref{3}) are added in order to make the expectational value of
each operator zero for the ground state $|g\rangle$:
\[
\langle g|H_{A}|g\rangle=\langle g|H_{B}|g\rangle=\langle g|V|g\rangle=0.
\]
Because the lowest eigenvalue of the total Hamiltonian $H$ is zero, $H$ is a
nonnegative operator:$~H\geq0$. Meanwhile, it should be noticed that $H_{B}$
has a negative eigenvalue, which can yield negative energy density. The ground
state is given by
\[
|g\rangle=\frac{1}{\sqrt{2}}\sqrt{1-\frac{h}{\sqrt{h^{2}+k^{2}}}}|+\rangle
_{A}|+\rangle_{B}-\frac{1}{\sqrt{2}}\sqrt{1+\frac{h}{\sqrt{h^{2}+k^{2}}}%
}|-\rangle_{A}|-\rangle_{B},
\]
where $|\pm\rangle_{A}~\left(  |\pm\rangle_{B}\right)  $ is the eigenstate of
$\sigma_{A}^{z}~\left(  \sigma_{B}^{z}\right)  $ with eigenvalue $\pm1$.

When local operations are performed to this system, we connect the system with
a local appratus and (or) add local external forces like  magnetic field.
However, time interval of the operations is assumed to be quite short, as
often argued in quantum information theory. Thus interaction terms between the
system and the external apparatus vanish in the Hamiltonian after the
short-time operation. 

Let $S_{M_{A}}$ denote a set of POVM measurements \cite{nc} for A which
measurement operators $M_{A}(\mu)$ with measurement output $\mu$ commute with
the interaction Hamiltonian $V$. The measurement operator $M_{A}(\mu)$ takes
the form of
\[
M_{A}(\mu)=e^{i\delta_{\mu}}\left(  m_{\mu}+e^{i\alpha_{\mu}}l_{\mu}\sigma
_{A}^{x}\right)  .
\]
The coefficients $m_{\mu}$, $l_{\mu}$, $\alpha_{\mu}$ and $\delta_{\mu}$ are
real constants which satisfy
\begin{align*}
\sum_{\mu}\left(  m_{\mu}^{2}+l_{\mu}^{2}\right)   &  =1,\\
\sum_{\mu}m_{\mu}l_{\mu}\cos\alpha_{\mu}  &  =0.
\end{align*}
The POVM corresponding to $M_{A}(\mu)$ is defined by
\[
\Pi_{A}(\mu)=M_{A}(\mu)^{\dag}M_{A}(\mu),
\]
which satisfies the completeness relation,
\[
\sum_{\mu}\Pi_{A}(\mu)=1_{A}.
\]
By introducing the emergence probability $p_{A}(\mu)$ of output $\mu$ for the
ground state and a real parameter $q_{A}(\mu)$, the POVM is written as
follows:
\[
\Pi_{A}(\mu)=p_{A}(\mu)+q_{A}(\mu)\sigma_{A}^{x}.
\]
By taking suitable values of $m_{\mu}$, $l_{\mu}$, and $\alpha_{\mu}$, all
values of $p_{A}(\mu)$ and $q_{A}(\mu)$ are permissible as long as they
satisfy $\sum_{\mu}p_{A}(\mu)=1$, $\sum_{\mu}q_{A}(\mu)=0$ and $p_{A}(\mu
)\geq\left\vert q_{A}(\mu)\right\vert $. The post-measurement state of the two
qubits with output $\mu$ is given by
\[
|A(\mu)\rangle=\frac{1}{\sqrt{p_{A}(\mu)}}M_{A}(\mu)|g\rangle.
\]
This measurement excites the system and the average post-measurement state has
a positive expectational value $E_{A}$ of $H$, which energy distribution is
localized at A. In fact, the value defined by
\[
E_{A}=\sum_{\mu}\langle g|M_{A}(\mu)^{\dag}HM_{A}(\mu)|g\rangle
\]
is computed as
\begin{equation}
E_{A}=\sum_{\mu}\langle g|M_{A}(\mu)^{\dag}H_{A}M_{A}(\mu)|g\rangle
=\frac{2h^{2}}{\sqrt{h^{2}+k^{2}}}\sum_{\mu}l_{\mu}^{2}. \label{4}%
\end{equation}
This infused energy $E_{A}$ is regarded as the QET energy input via the
measurement of A. During the measurement, $E_{A}$ is transfered from external
systems including the measurement device with a battery respecting local
energy conservation. The QET energy conservation law during local measurements
to a qubit of a spin chain has been discussed in \cite{hotta1}.

The key feature of this model is that any measurement of $S_{M_{A}}$ does not
increase the average energy of B at all. By explicit calculations, the average
values of the Hamiltonian contributions $H_{B}$ and $V$ are found to remain
zero after the measurement and are the same as those of the ground state. This
measurement does not yield instantaneous change of $V$. Therefore we have no
direct force of A affecting B after the measurement. Thus, we cannot extract
energy from B by the standard way soon after the measurement. In fact, if any
local unitary operation $W_{B}$ independent of A's measurement result is
performed to B, the post-operation state $\omega$ is given by
\[
\omega=W_{B}\left(  \sum_{\mu}M_{A}(\mu)|g\rangle\langle g|M_{A}(\mu)^{\dag
}\right)  W_{B}^{\dag}.
\]
The energy difference after the operation is calculated as
\begin{equation}
\operatorname*{Tr}\left[  \omega H\right]  -E_{A}=\langle g|W_{B}^{\dag
}\left(  H_{B}+V\right)  W_{B}|g\rangle,\label{e01}%
\end{equation}
where we have used
\[
W_{B}^{\dag}H_{A}W_{B}=H_{A}W_{B}^{\dag}W_{B}=H_{A},
\]%
\[
\left[  W_{B}^{\dag}\left(  H_{B}+V\right)  W_{B},~M_{A}(\mu)\right]  =0,
\]
and the completeness relation of the POVM's:%
\[
\sum_{\mu}M_{A}(\mu)^{\dag}M_{A}(\mu)=1_{A}.
\]
From Eq. (\ref{e01}), it is proven that the energy difference is nonnegative:
\[
\operatorname*{Tr}\left[  \omega H\right]  -E_{A}=\langle g|W_{B}^{\dag}%
HW_{B}|g\rangle\geq0,
\]
because of a relation such that $\langle g|W_{B}^{\dag}H_{A}W_{B}%
|g\rangle=\langle g|H_{A}|g\rangle=0$ and the nonnegativity of $H$. Therefore,
as a natural result, no local operation to B independent of $\mu$ extracts
energy from the system.

After a while, the infused energy $E_{A}$ diffuses to B. The time evolution of
the expectational values $H_{B}$ and $V$ of the average post-measurement state
is calculated as%
\[
\langle H_{B}(t)\rangle=\sum_{\mu}p_{A}(\mu)\langle A(\mu)|e^{itH}%
H_{B}e^{-itH}|A(\mu)\rangle=\frac{h^{2}\sum_{\mu}l_{\mu}^{2}}{\sqrt
{h^{2}+k^{2}}}\left[  1-\cos\left(  4kt\right)  \right]  ,
\]
and $\langle V(t)\rangle=0$. Therefore, energy can be extracted from B after a
diffusion time scale of $1/k$; this is just a usual energy transportation from
A to B. Amazingly, the QET protocol can transport energy from A to B in a time
scale much shorter than that of the usual transportation. In the \ protocol,
the measurement output $\mu$ is announced to B. Because the model is
non-relativistic, the propagation speed of the announced output can be much
faster than the diffusion speed of the infused energy and can be approximated
as infinity. Soon after the arrival of the output $\mu$, we perform a local
operation $U_{B}(\mu)$ on B dependent on $\mu$. Then, the average state after
the operation is given by%

\[
\rho=\sum_{\mu}U_{B}(\mu)M_{A}(\mu)|g\rangle\langle g|M_{A}(\mu)^{\dag}%
U_{B}(\mu)^{\dag}.
\]
In Figure 1, a schematic diagram of this QET model is presented. The
expectational value of the total energy after the operation is given by
\[
\operatorname*{Tr}\left[  \rho H\right]  =\sum_{\mu}\langle g|M_{A}(\mu
)^{\dag}U_{B}(\mu)^{\dag}HU_{B}(\mu)M_{A}(\mu)|g\rangle.
\]
On the basis of the fact that $U_{B}(\mu)$ commutes with $H_{A}$ and Eq.
(\ref{4}), $E_{B}$ is computed as
\[
E_{B}=E_{A}-\operatorname*{Tr}\left[  \rho H\right]  =-\operatorname*{Tr}%
\left[  \rho\left(  H_{B}+V\right)  \right]  .
\]
Further, on the basis of the fact that $M_{A}(\mu)$ commutes with $U_{B}(\mu
)$, $H_{B}$ and $V$, the energy can be written as
\[
E_{B}=-\sum_{\mu}\langle g|\Pi_{A}(\mu)\left(  H_{B}(\mu)+V(\mu)\right)
|g\rangle,
\]
where the $\mu$-dependent operators are given by $H_{B}(\mu)=U_{B}(\mu)^{\dag
}H_{B}U_{B}(\mu)$ and $V(\mu)=U_{B}(\mu)^{\dag}VU_{B}(\mu)$. Here, let us
write the general form of $U_{B}(\mu)$ as follows:
\[
U_{B}(\mu)=\cos\omega_{\mu}+i\vec{n}_{\mu}\cdot\vec{\sigma}_{B}\sin\omega
_{\mu},
\]
where $\omega_{\mu}\,$\ is a real parameter, $\vec{n}_{\mu}=\left(  n_{x\mu
},n_{y\mu},n_{z\mu}\right)  \,$is a three-dimensional unit real vector and
$\vec{\sigma}_{B}$ is the Pauli spin vector operator of B. Then, an explicit
evaluation of $E_{B}$ becomes possible. The result is expressed as
\[
E_{B}=\frac{1}{\sqrt{h^{2}+k^{2}}}\sum_{\mu}Q(\mu),
\]
where $Q(\mu)$ is given by
\begin{equation}
Q(\mu)=X(\mu)\cos\left(  2\omega_{\mu}\right)  -hkq_{A}(\mu)n_{y\mu}%
\sin(2\omega_{\mu})-X(\mu), \label{6}%
\end{equation}
where $X(\mu)$ is defined by
\[
X(\mu)=p_{A}(\mu)\left[  h^{2}\left(  1-n_{z\mu}^{2}\right)  +2k^{2}\left(
1-n_{x\mu}^{2}\right)  \right]  -3hkq_{A}(\mu)n_{x\mu}n_{z\mu}.
\]
In order to maximize the teleported energy $E_{B}$ for a given POVM
measurement of A, let us first maximize $Q(\mu)$ in Eq. (\ref{6}) by changing
the parameter $\omega_{\mu}$. This maximum value is calculated as%

\begin{equation}
\max_{\omega_{\mu}}Q(\mu)=\sqrt{X(\mu)^{2}+\left[  hkq_{A}(\mu)n_{y\mu
}\right]  ^{2}}-X(\mu). \label{13}%
\end{equation}
Next, let us introduce a parametrization of $n_{x\mu}$ and $n_{z\mu}$ as
$n_{x\mu}=\sqrt{z}\cos\psi_{\mu}$ and $n_{z\mu}=\sqrt{z}\sin\psi_{\mu}$ for
fixed $z=1-n_{y\mu}^{2}$ which runs over $\left[  0,1\right]  $, where
$\psi_{\mu}$ is a real parameter. It is observed that $\max_{\omega_{\mu}%
}Q(\mu)$ in Eq. (\ref{13}) is a monotonically decreasing function of $X(\mu)$.
Thus, we must find the minimum value of $X(\mu)$ in terms of $\psi_{\mu}$. By
using the parametrization, we can minimize $X(\mu)$ as%

\[
\min_{\psi_{\mu}}X(\mu)=\left(  1-\frac{z}{2}\right)  p_{A}(\mu)\left(
h^{2}+2k^{2}\right)  -\frac{z}{2}\sqrt{\left(  h^{2}-2k^{2}\right)  ^{2}%
p_{A}(\mu)^{2}+9h^{2}k^{2}q_{A}(\mu)^{2}}.
\]
Therefore, the maximum value of $\max_{\omega_{\mu}}Q(\mu)$ in terms of
$\psi_{\mu}$ is obtained as follows:%

\[
\max_{\omega_{\mu},\psi(\mu)}Q(\mu)=\sqrt{\left(  \min_{\psi(\mu)}%
X(\mu)\right)  ^{2}+h^{2}k^{2}q_{A}(\mu)^{2}(1-z)}-\min_{\psi(\mu)}X(\mu).
\]
Next, in order to maximize $\max_{\omega_{\mu},\psi_{\mu}}Q(\mu)$ in terms of
$z$, let us write it as a function $T(z)$ of $z$:
\[
T(z)=\max_{\omega_{\mu},\psi_{\mu}}Q(\mu)=\sqrt{\left(  a-bz\right)
^{2}+c\left(  1-z\right)  }-\left(  a-bz\right)  ,
\]
where $a$, $b$ and $c$ are positive constants given by%
\begin{align*}
a  &  =p_{A}(\mu)\left(  h^{2}+2k^{2}\right)  ,\\
b  &  =\frac{p_{A}(\mu)}{2}\left(  h^{2}+2k^{2}\right)  +\frac{1}{2}%
\sqrt{\left(  h^{2}-2k^{2}\right)  ^{2}p_{A}(\mu)^{2}+9h^{2}k^{2}q_{A}%
(\mu)^{2}},\\
c  &  =h^{2}k^{2}q_{A}(\mu)^{2}.
\end{align*}
The derivative of $T(z)$ can be calculated as
\[
\partial_{z}T(z)=\frac{t(z)}{2\sqrt{\left(  a-bz\right)  ^{2}+c\left(
1-z\right)  }},
\]
where $t(z)$ is a function given by
\[
t(z)=-c+2b\left(  \sqrt{\left(  a-bz\right)  ^{2}+c\left(  1-z\right)
}-\left(  a-bz\right)  \right)  .
\]
It can be verified that $t(z)$ and $\partial_{z}T(z)$ are nonpositive for
$z\in\left[  0,1\right]  $. This verification can be done as follows. Let us
first consider an equation $t(\bar{z})=0$. It turns out that, in the
transformation of this equation for solving $\bar{z}$, the dependence of
$\bar{z}$ gets lost and we get just a constraint condition on $p_{A}(\mu)$ and
$q_{A}(\mu)$ such that $q_{A}(\mu)^{2}\left(  p_{A}(\mu)^{2}-q_{A}(\mu
)^{2}\right)  =0$. Thus, if $p_{A}(\mu)^{2}=q_{A}(\mu)^{2}$ or $q_{A}(\mu
)^{2}=0$, the equation $t(z)=0$ holds for all $z\in\left[  0,1\right]  $. If
$p_{A}(\mu)^{2}\neq q_{A}(\mu)^{2}$ and $q_{A}(\mu)^{2}\neq0$, the solution
$\bar{z}$ does not exist and $t(z)$ has a definite sign for $z\in\left[
0,1\right]  $. In order to check the sign, let us substitute $z=1$ into
$t(z)$. Then, when $a\geq b$, we get $t\left(  1\right)  =-h^{2}k^{2}q_{A}%
(\mu)^{2}<0$, and when $a\leq b$, $t\left(  1\right)  =-8h^{2}k^{2}(p_{A}%
(\mu)^{2}-q_{A}(\mu)^{2})<0$. Thus, it is verified that $t(z)$\ and
$\partial_{z}T(z)$ are nonpositive. Therefore, $T(z)$ takes the maximum value
at $z=0$. This implies that $Q(\mu)$ can be maximized as $\max_{U_{B}(\mu
)}Q(\mu):=\max_{\omega_{\mu},\psi_{\mu},z}Q(\mu)=T(0)$. This leads to our
final expression of the maximum teleported energy for the measurement, which
is clearly nonnegative, as follows:
\begin{equation}
\max_{U_{B}(\mu)}E_{B}=\frac{h^{2}+2k^{2}}{\sqrt{h^{2}+k^{2}}}\sum_{\mu}%
p_{A}(\mu)\left[  \sqrt{1+\frac{h^{2}k^{2}}{\left(  h^{2}+2k^{2}\right)  ^{2}%
}\frac{q_{A}(\mu)^{2}}{p_{A}(\mu)^{2}}}-1\right]  . \label{21}%
\end{equation}
The operation $U_{\max}(\mu)$ which attains the maximum of teleported energy
is given by
\[
U_{\max}(\mu)=\cos\Omega_{\mu}+i\sigma_{B}^{y}\sin\Omega_{\mu},
\]
where $\Omega_{\mu}$ is a real constant which satisfies%

\begin{align*}
\cos\left(  2\Omega_{\mu}\right)   &  =\frac{\left(  h^{2}+2k^{2}\right)
p_{A}(\mu)}{\sqrt{\left(  h^{2}+2k^{2}\right)  ^{2}p_{A}(\mu)^{2}+h^{2}%
k^{2}q_{A}(\mu)^{2}}},\\
\sin\left(  2\Omega_{\mu}\right)   &  =-\frac{hkq_{A}(\mu)}{\sqrt{\left(
h^{2}+2k^{2}\right)  ^{2}p_{A}(\mu)^{2}+h^{2}k^{2}q_{A}(\mu)^{2}}}.
\end{align*}
Besides, the teleported energy can be maximized among POVM measurements of
$S_{M_{A}}$. This is achieved when each POVMs are proportional to projective
operators and given by%
\[
\max_{S_{M_{A}},U_{B}(\mu)}E_{B}=\frac{h^{2}+2k^{2}}{\sqrt{h^{2}+k^{2}}%
}\left[  \sqrt{1+\frac{h^{2}k^{2}}{\left(  h^{2}+2k^{2}\right)  ^{2}}%
}-1\right]  .
\]

\bigskip

\section{Relation between Entanglement Breaking and Teleported Energy}

\bigskip

~

Next, we analyze entanglement breaking by the POVM measurement of A and show
two inequalities between the maximum teleported energy and the entanglement
breaking. We adopt entropy of entanglement as a quantitative measure of
entanglement. The entropy of a pure state $|\Psi_{AB}\rangle$ of A and B is
defined as
\[
S_{AB}=-\operatorname*{Tr}_{B}\left[  \operatorname*{Tr}_{A}\left[  |\Psi
_{AB}\rangle\langle\Psi_{AB}|\right]  \ln\operatorname*{Tr}_{A}\left[
|\Psi_{AB}\rangle\langle\Psi_{AB}|\right]  \right]  .
\]
Before the measurement, the total system is prepared to be in the ground state
$|g\rangle$. \ The reduced state of B is given by
\[
\rho_{B}=\operatorname*{Tr}_{A}\left[  |g\rangle\langle g|\right]  .
\]
After the POVM measurement outputting $\mu$, the state is transferred into a
pure state $|A(\mu)\rangle$. The reduced post-measurement state of B is
calculated as
\[
\rho_{B}(\mu)=\frac{1}{p_{A}(\mu)}\operatorname*{Tr}_{A}\left[  \Pi_{A}%
(\mu)|g\rangle\langle g|\right]  .
\]
The entropy of entanglement of the ground state is given by
\[
S_{AB}(g)=-\operatorname*{Tr}_{B}\left[  \rho_{B}\ln\rho_{B}\right]
\]
and that of the post-measurement state with output $\mu$ is given by
\[
S_{AB}(\mu)=-\operatorname*{Tr}_{B}\left[  \rho_{B}(\mu)\ln\rho_{B}%
(\mu)\right]  .
\]
By using these results, we define the consumption of ground-state entanglement
by the measurement as the difference between the ground-state entanglement and
the averaged post-measurement-state entanglement:%

\[
\Delta S_{AB}=S_{AB}(g)-\sum_{\mu}p_{A}(\mu)S_{AB}(\mu).
\]
Interestingly, this quantity is tied to the quantum mutual information between
the measurement result of A and the post-measurement state of B. Let us
introduce a Hilbert space for a measurement pointer system $\bar{A}$ of the
POVM measurement, which is spanned by orthonormal states $|\mu_{\bar{A}%
}\rangle$ corresponding to the output $\mu$ satisfying $\langle\mu_{\bar{A}%
}|\mu_{\bar{A}}^{\prime}\rangle=\delta_{\mu\mu^{\prime}}$. Then, the average
state of $\bar{A}$ and B after the measurement is given by
\[
\Phi_{\bar{A}B}=\sum_{\mu}p_{A}(\mu)|\mu_{\bar{A}}\rangle\langle\mu_{\bar{A}%
}|\otimes\rho_{B}(\mu).
\]
By using the reduced operators $\Phi_{\bar{A}}=\operatorname*{Tr}_{B}\left[
\Phi_{\bar{A}B}\right]  $ and $\Phi_{B}=\operatorname*{Tr}_{\bar{A}}\left[
\Phi_{\bar{A}B}\right]  $, the mutual information $I_{\bar{A}B}$ is defined as%
\begin{equation}
I_{\bar{A}B}=-\operatorname*{Tr}_{\bar{A}}\left[  \Phi_{\bar{A}}\ln\Phi
_{\bar{A}}\right]  -\operatorname*{Tr}_{B}\left[  \Phi_{B}\ln\Phi_{B}\right]
+\operatorname*{Tr}_{\bar{A}B}\left[  \Phi_{\bar{A}B}\ln\Phi_{\bar{A}%
B}\right]  .\nonumber
\end{equation}
By using $\operatorname*{Tr}_{B}\left[  \Phi_{\bar{A}B}\right]  =\sum_{\mu
}p_{A}(\mu)|\mu_{\bar{A}}\rangle\langle\mu_{\bar{A}}|$ and $\operatorname*{Tr}%
_{\bar{A}}\left[  \Phi_{\bar{A}B}\right]  =\sum_{\mu}p_{A}(\mu)\rho_{B}%
(\mu)=\rho_{B}$, it can be straightforwardly proven that $I_{\bar{A}B}$ is
equal to $\Delta S_{AB}$. This relation provides another physical
interpretation of $\Delta S_{AB}$.

Next, let us calculate $\Delta S_{AB}$ explicitly. All the eigenvalues of
$\rho_{B}(\mu)$ are given by%

\begin{equation}
\lambda_{\pm}(\mu)=\frac{1}{2}\left[  1\pm\sqrt{\cos^{2}\varsigma+\sin
^{2}\varsigma\frac{q_{A}(\mu)^{2}}{p_{A}(\mu)^{2}}}\right]  , \label{29}%
\end{equation}
where $\varsigma$ is a real constant which satisfies
\[
\cos\varsigma=\frac{h}{\sqrt{h^{2}+k^{2}}},~\sin\varsigma=\frac{k}{\sqrt
{h^{2}+k^{2}}}.
\]
The eigenvalues of $\rho_{B}$ are obtained by substituting $q_{A}(\mu)=0$ into
Eq. (\ref{29}). By using $\lambda_{s}(\mu)$, $\Delta S_{AB}$ can be evaluated
as
\begin{equation}
\Delta S_{AB}=\sum_{\mu}p_{A}(\mu)f_{I}\left(  \frac{q_{A}(\mu)^{2}}{p_{A}%
(\mu)^{2}}\right)  , \label{30}%
\end{equation}
where $f_{I}(x)$ is a monotonically increasing function of $x\in\left[
0,1\right]  $ and is defined by%
\begin{align*}
f_{I}(x)  &  =\frac{1}{2}\left(  1+\sqrt{\cos^{2}\varsigma+x\sin^{2}\varsigma
}\right)  \ln\left(  \frac{1}{2}\left(  1+\sqrt{\cos^{2}\varsigma+x\sin
^{2}\varsigma}\right)  \right) \\
&  +\frac{1}{2}\left(  1-\sqrt{\cos^{2}\varsigma+x\sin^{2}\varsigma}\right)
\ln\left(  \frac{1}{2}\left(  1-\sqrt{\cos^{2}\varsigma+x\sin^{2}\varsigma
}\right)  \right) \\
&  -\frac{1}{2}\left(  1+\cos\varsigma\right)  \ln\left(  \frac{1}{2}\left(
1+\cos\varsigma\right)  \right) \\
&  -\frac{1}{2}\left(  1-\cos\varsigma\right)  \ln\left(  \frac{1}{2}\left(
1-\cos\varsigma\right)  \right)  .
\end{align*}
It is worth noting that the optimal teleported energy $\max_{U_{B}(\mu)}E_{B}$
in Eq. (\ref{21}) takes a form similar to Eq. (\ref{30}) as%

\begin{equation}
\max_{U_{B}(\mu)}E_{B}=\sum_{\mu}p_{A}(\mu)f_{E}\left(  \frac{q_{A}(\mu)^{2}%
}{p_{A}(\mu)^{2}}\right)  , \label{003}%
\end{equation}
where $f_{E}(x)$ is a monotonically increasing function of $x\in\left[
0,1\right]  $ and is defined by
\[
f_{E}(x)=\sqrt{h^{2}+k^{2}}\left(  1+\sin^{2}\varsigma\right)  \left[
\sqrt{1+\frac{\cos^{2}\varsigma\sin^{2}\varsigma}{\left(  1+\sin^{2}%
\varsigma\right)  ^{2}}x}-1\right]  .
\]
Expanding both $f_{I}(x)$ and $f_{E}(x)$ around $x=0$ yields%

\begin{align*}
f_{I}(x)  &  =\frac{\sin^{2}\varsigma}{4\cos\varsigma}\ln\frac{1+\cos
\varsigma}{1-\cos\varsigma}x+O(x^{2}),\\
f_{E}(x)  &  =\sqrt{h^{2}+k^{2}}\frac{\cos^{2}\varsigma\sin^{2}\varsigma
}{2\left(  1+\sin^{2}\varsigma\right)  }x+O(x^{2}).
\end{align*}
By deleting $x$ in the above two equations, we obtain the following relation
for weak measurements with infinitesimally small $q_{A}(\mu)$:%

\[
\Delta S_{AB}=\frac{1+\sin^{2}\varsigma}{2\cos^{3}\varsigma}\ln\frac
{1+\cos\varsigma}{1-\cos\varsigma}\frac{\max_{U_{B}(\mu)}E_{B}}{\sqrt
{h^{2}+k^{2}}}+O(q_{A}(\mu)^{4}).
\]
It is of great significance that this relation can be extended as the
following inequality for general measurements of $S_{M_{A}}:$%

\begin{equation}
\Delta S_{AB}\geq\frac{1+\sin^{2}\varsigma}{2\cos^{3}\varsigma}\ln\frac
{1+\cos\varsigma}{1-\cos\varsigma}\frac{\max_{U_{B}(\mu)}E_{B}}{\sqrt
{h^{2}+k^{2}}}. \label{32}%
\end{equation}
This inequality is one of the main results of this Letter and implies that a
large amount of teleported energy really requests a large amount of
consumption of the ground-state entanglement between A and B. The proof of Eq.
(\ref{32}) is as follows. Let us introduce two rescaled functions as follows.
\begin{align*}
\bar{f}_{I}(x)  &  =4\frac{\cos\varsigma}{\sin^{2}\varsigma}\left(  \ln
\frac{1+\cos\varsigma}{1-\cos\varsigma}\right)  ^{-1}f_{I}(x),\\
\bar{f}_{E}(x)  &  =\frac{1}{\sqrt{h^{2}+k^{2}}}\frac{2\left(  1+\sin
^{2}\varsigma\right)  }{\cos^{2}\varsigma\sin^{2}\varsigma}f_{E}(x).
\end{align*}
It can be easily shown that $\bar{f}_{E}(x)$ is a convex function:
\[
\partial_{x}^{2}\bar{f}_{E}(x)<0.
\]
From this convexity and $\bar{f}_{E}(x)=x+O(x^{2})$, the function $\bar{f}%
_{E}(x)$ satisfies
\begin{equation}
\bar{f}_{E}(x)\leq x \label{46}%
\end{equation}
for $0\leq x\leq1$. On the other hand, it is observed that $\bar{f}_{I}(x)$ is
a concave function, as shown below. The derivative of $\bar{f}_{I}(x)$ is
computed as
\[
\partial_{x}\bar{f}_{I}(x)=\frac{2\cos\varsigma}{\ln\frac{1+\cos\varsigma
}{1-\cos\varsigma}}g_{I}(y(x)),
\]
where $y(x)=\sqrt{\cos^{2}\varsigma+x\sin^{2}\varsigma}$ and $g_{I}(y)$ is a
positive function of $y$ defined as
\[
g_{I}(y)=\frac{1}{y}\ln\frac{1+y}{1-y}.
\]
It should be noted that $y(x)$ is a monotonically increasing function of $x$.
The derivative of $g_{I}(y)$ is calculated as
\[
\partial_{y}g_{I}(y)=\frac{s_{I}(y)}{y^{2}},
\]
where $s_{I}(y)$ is a function of $y$ given by
\[
s_{I}(y)=\frac{2y}{1-y^{2}}-\ln\left(  \frac{1+y}{1-y}\right)  ,
\]
and satisfies a boundary condition as $s_{I}(0)=0$. It is also easy to show
that the derivative of $s_{I}(y)$ is positive for $y>0$:
\[
\partial_{y}s_{I}(y)=\frac{4y^{2}}{\left(  1-y^{2}\right)  ^{2}}>0.
\]
Thus, $s_{I}(y)$ and $\partial_{y}g_{I}(y)$ are positive for $y\in\left[
\cos\varsigma,1\right]  $. From these results, it has been proven that
$\bar{f}_{I}(x)$ is concave for $x\in\left[  0,1\right]  $: $\partial_{x}%
^{2}\bar{f}_{I}(x)>0$. From this concavity and $\bar{f}_{I}(x)=x+O(x^{2})$, it
is shown that $\bar{f}_{I}(x)$ satisfies the relation
\begin{equation}
\bar{f}_{I}(x)\geq x. \label{45}%
\end{equation}
Because of Eqs. (\ref{46}) and (\ref{45}), we obtain the following inequality:%

\[
\bar{f}_{I}(x)\geq x\geq\bar{f}_{E}(x).
\]
This implies that the energy-entanglement inequality in Eq. (\ref{32}) holds.
In addition, we can prove another inequality between energy and entanglement
breaking. Because the convex function $\bar{f}_{E}(x)$ and the concave
function $\bar{f}_{I}(x)$ are monotonically increasing functions of
$x\in\left[  0,1\right]  $ which satisfy $\bar{f}_{E}(0)=$ $\bar{f}_{I}(0)=0$,
we have the following relation:%
\[
\bar{f}_{E}(x)\geq\frac{\bar{f}_{E}(1)}{\bar{f}_{I}(1)}\bar{f}_{I}(x).
\]
Consequently, the following inequality, which is another main result of this
Letter, is obtained for all measurements of $S_{M_{A}}$:
\begin{equation}
\max_{U_{B}(\mu)}E_{B}\geq\frac{2\sqrt{h^{2}+k^{2}}\left[  \sqrt{4-3\cos
^{2}\varsigma}-2+\cos^{2}\varsigma\right]  }{\left(  1+\cos\varsigma\right)
\ln\left(  \frac{2}{1+\cos\varsigma}\right)  +\left(  1-\cos\varsigma\right)
\ln\left(  \frac{2}{1-\cos\varsigma}\right)  }\Delta S_{AB}. \label{770}%
\end{equation}
This ensures that if we have consumption of ground-state entanglement $\Delta
S_{AB}$ for a measurement of $S_{M_{A}}$, we can in principle teleport energy
from A to B, where the energy amount is greater than the value of the
right-hand-side term of Eq. (\ref{770}). This bound is achieved for non-zero
energy transfer by measurements with $q_{A}(\mu)=\pm p_{A}(\mu)$. The
inequalities in Eq. (\ref{32}) and Eq. (\ref{770}) help us to gain a deeper
understanding of entanglement as a physical resource because they show that
the entanglement decrease by the measurement of A is directly related to the
increase of the available energy at B as an evident physical resource.

\bigskip

\textbf{Acknowledgments}

I would like to thank Yasusada Nambu for a comment. This research has been
partially supported by the Global COE Program of MEXT, Japan, and the Ministry
of Education, Science, Sports and Culture, Japan, under Grant No. 21244007.

\bigskip

Figure Caption

\bigskip

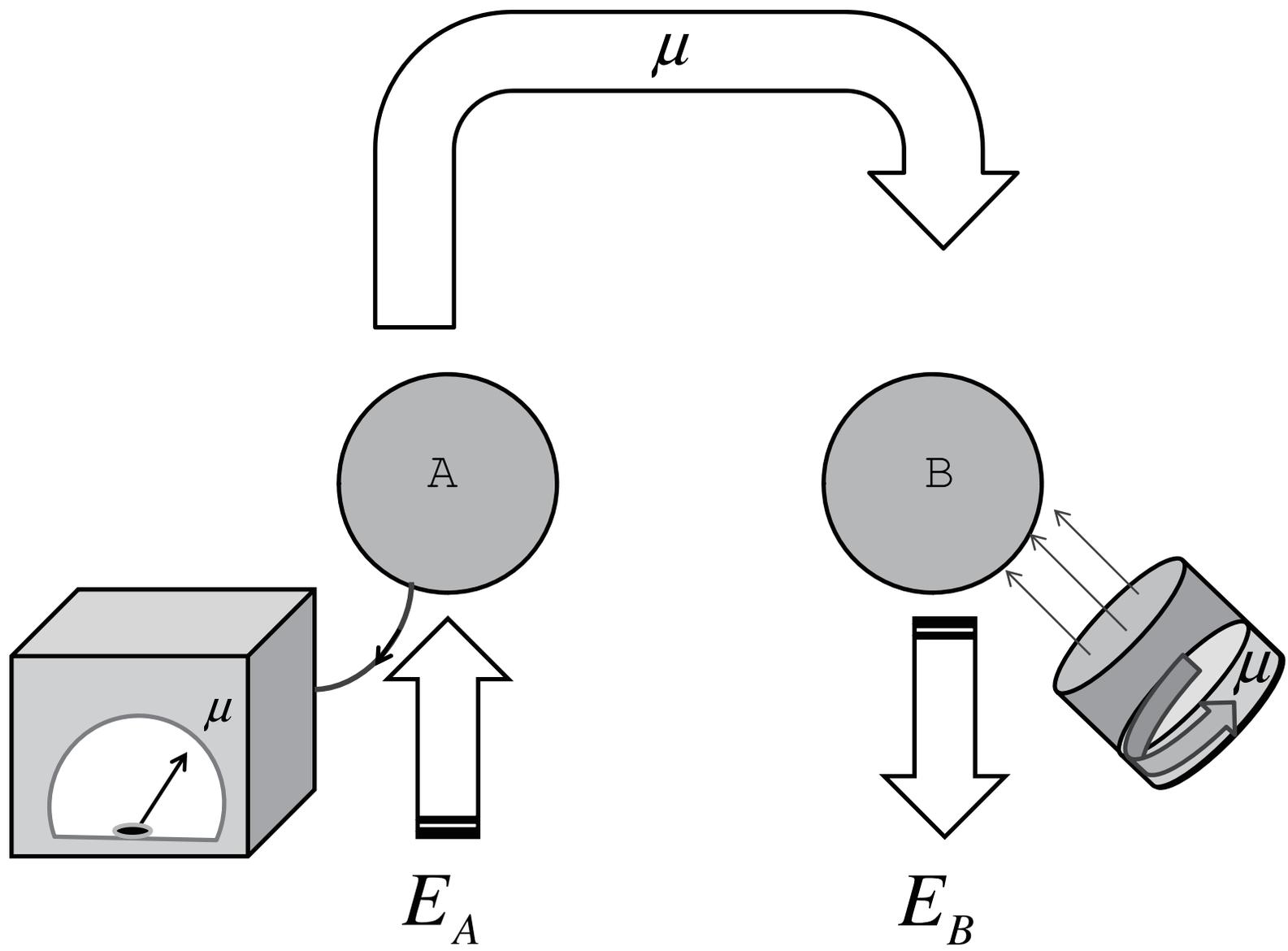
Figure 1: Schematic diagram of the minimal QET model. A POVM measurement is
performed on A with infusion of energy $E_{A}$. The measurement result $\mu$
is announced to B through a classical channel. After the arrival of $\mu$, a
unitary operation dependent on $\mu$ is performed on B with extraction of
energy $E_{B}$.

\bigskip
\end{document}